\documentstyle[12pt,times,graphicx,subfigure,pstricks,amssymb]{article}

\large \normalsize
\def \be {\begin{equation}}
\def \ee {\end{equation}}
\def\norm#1{\left \vert#1\right \vert}

\title{First numerical evidence of global Arnold diffusion in quasi--integrable 
systems}

 \author{ Massimiliano Guzzo\footnote{Dipartimento di Matematica Pura 
ed Applicata, Universit\`a degli Studi di Padova, via Belzoni 7, 35131 Padova,
Italy.}, 
Elena Lega,  Claude Froeschl\'e\footnote{Observatoire de Nice, Bv. de l'Observatoire, B.P.~4229, 06304 Nice 
cedex 4, France.}}

\begin{document}
\maketitle
\begin{abstract}
We provide numerical evidence of global diffusion occurring in slightly 
perturbed integrable Hamiltonian systems and symplectic maps. We show that 
even if a system is sufficiently close to be integrable, global diffusion 
occurs on a set with peculiar topology, the so--called Arnold web, and is 
qualitatively different from Chirikov diffusion, occurring in more 
perturbed systems.
\end{abstract}  

\section{Introduction}
The characterization of mechanisms for diffusion of orbits in 
quasi--integrable Hamiltonian systems and symplectic maps is a relevant 
topic for many fields of physics, such as celestial mechanics, 
dynamical astronomy, statistical physics, plasma physics and 
particle accelerators. Any dynamical state of an integrable system can 
be characterized with a set of action--angle conjugate variables 
$(I_1,\ldots ,I_n, \phi_1,\ldots ,\phi_n)$, where the actions $I_j$ 
are constants of motion, while the angles  $\phi_j$   
simply change linearly with time. The properties of the system which are 
relevant for the stability are determined by the actions. 

Small perturbations of integrable systems can produce a slow drift of the 
initial value of the actions, and after certain time these   
drifts can cumulate in such a way to drive the system into very 
different physical state. By global drift we mean that orbits explore 
macroscopic regions of the action--space. If the drift 
behaves as a diffusion process  (precisely, we mean that the numerically 
computed mean squared distance of $I_1,\ldots ,I_n$ with respect to their 
initial values grows, on average,  linearly with time), we will refer to it 
as global diffusion. 

In 1979 Chirikov [1] described a possible model for global drift valid 
when the perturbation is greater than some critical value.  
Chirikov's model has so far been successfully used to describe diffusion in 
systems from different fields of physics (see, for example, [2]). 
One of the reasons of the broad detection 
of the  Chirikov's diffusion is that its typical times fall within the simulation abilities of modern computers as far back as the seventies. 

For smaller perturbations the 
systems fall within the range of celebrated perturbation theories such 
as KAM [3,4,5] and Nekhoroshev theorems [6], which leave the 
possibility for global drift only on a subset of the possible dynamical 
states with peculiar topology, the so--called Arnold web, and force 
diffusion times to be at least exponentially long with an inverse power of 
the norm of the perturbation.  
The theoretical possibility of global drift in quasi--integrable 
systems has been first shown in 1964 by Arnold [7] for a specific 
system, and is commonly called Arnold diffusion. Recently, Arnold's result 
has been much generalized [8]. Being interested to applications to 
specific systems, and in particular to systems of interest for physics, 
we use a numerical approach which, avoiding theoretical difficulties, 
measures directly the quantitative features of eventual long term diffusion. 
In 2003 we numerically detected a very slow local diffusion confined to the 
Arnold web [9] in a model perturbed system. In this paper we provide 
numerical evidence both on quasi--integrable 
Hamiltonian systems and symplectic maps of a relevant phenomenon of global 
diffusion of orbits occurring on the Arnold web. More precisely, we show  
that a set of well chosen initial 
conditions practically explores the whole web and the process behaves as a 
global diffusion.

In section 2 we will describe the Arnold web of quasi--integrable Hamiltonian 
systems and symplectic maps and in section 3 we will describe the numerical 
methods used to detect the Arnold web and the global diffusion on it.

\section{Arnold web of quasi--integrable systems}
For definiteness, we refer to the Hamiltonian system with 
Hamilton function:
\begin{equation}
 H_\epsilon = {I_1^2 \over 2}+{I_2^2 \over 2} +I_3 + \epsilon \ f\ \ \ \ ,
\ \ \ \  f\ =\ 
{ 1 \over \cos (\phi _1) +
 \cos (\phi _2) + \cos (\phi_3) + 3+c}\ \ ,
\label{hamilt}
\end{equation}
where $\epsilon$ is a small parameter and $c>0$, so that the equations of 
motion for $I_1,I_2,I_3\in {\Bbb R}$ and $\phi_1,\phi_2,\phi_3 \in 
{\Bbb S}^1$ are: $\dot I_i=-{\partial H_\epsilon\over \partial \phi_i}$ 
and $\dot \phi_i={\partial H_\epsilon\over \partial I_i}$ for any $i=1,2, 3$. 
In the integrable case (when $\epsilon=0$) the actions 
are constants of motion while the angles $\phi_1(t)=\phi_1(0)+I_1 t$, 
$\phi_2(t)=\phi_2(0)+I_2 t$, $\phi_3(t)=\phi_3(0)+t$ rotate with 
frequencies $\omega_1=I_1$, $\omega_2=I_2$, $\omega_3=1$. Therefore, each 
couple of actions $I_1,I_2$ characterizes an invariant
torus ${\Bbb T}^3$. For any small $\epsilon$ different from zero,
$H_\epsilon$ is not expected to be integrable. However, if $\epsilon$ 
is sufficiently small, the KAM theorem applies ( $H_\epsilon$ is real 
analytic and $H_0$ is isoenergetically non--degenerate): for any invariant 
torus of the original system with Diophantine non--resonant 
frequencies\footnote{$\omega_1,\omega_2,\omega_3$ are Diophantine 
if there exist positive constants $\gamma$, $\tau$ such that 
$\vert \sum_i k_i \omega_i \vert > \gamma /\vert k \vert^\tau$, with 
$\vert k \vert =\sum_i \vert k_i\vert$, for all $k=(k_1,k_2 ,k_3)\in
{\Bbb Z}^3\backslash 0$. The Diophantine condition considered 
in KAM theorem requires $\tau>2$ (for 3 degrees of freedom systems) 
 and $\gamma$ which suitably rescales 
with $\epsilon$.} there exists an invariant torus in the perturbed system 
which is a small deformation of the unperturbed one. The complement of  
the set made of these invariant tori is called Arnold web, and in such a set, 
in principle, the motions can exhibit chaotic diffusion. Because Nekhoroshev 
theorem also applies to the system, any eventual chaotic diffusion will 
occur on very long times that grow at least exponentially with an inverse 
power of $\epsilon$.  Precisely, because $H_0$ is quasi--convex, 
the following estimates apply: 
$\vert I(t)-I(0)\vert \leq a \epsilon^{1/6}$, for any 
$\vert t\vert \leq b \exp (\epsilon_0/\epsilon )^{1/6}$  
with $a,b$ suitable positive constants (see [10] and also [16]).

To describe the topology of the Arnold web, it is convenient to refer to 
the subset of the phase space determined by the section 
$S=\{(I_1,I_2)\in {\Bbb R}^2, \phi_i=0, i=1,2,3\}$. From the KAM theorem, it 
follows that any invariant torus cuts the section $S$ in only one point
$(I_1,I_2)$. Moreover, the Diophantine condition on the frequencies implies 
that these points are outside a neighborhood of lines 
$k_1 I_1+k_2 I_2 +k_3=0$ proportional to $\gamma /\vert k \vert^\tau$, 
for any $k=(k_1,k_2,k_3) \in {\Bbb Z}^3\backslash 0$. 
Therefore, the intersection 
between the Arnold web and the section $S$ consists of all these straight 
lines with a neighborhood which decreases as $\vert k \vert$ increases. 
Any of these straight lines is called resonance, and the integer 
$\vert k \vert$ is called resonance order. The Arnold web is open, dense, and  
has a small relative measure (proportional to $\gamma$).

The case of quasi--integrable symplectic maps is analogous. In this paper we 
consider the following quasi--integrable symplectic map:
\begin{eqnarray}
\phi_1'=\phi_1+I_1\ \ &,&\ \ \phi_2'=\phi_2+I_2\cr
I'_1=I_1+\epsilon {\partial f\over \partial \phi_1}(\phi_1+I_1,
\phi_2+I_2)\ \ &,&\ \ \ 
I'_2=I_2+\epsilon {\partial f\over \partial \phi_2}(\phi_1+I_1,
\phi_2+I_2)
\end{eqnarray}
where $f=1/(\cos(\phi_1)+\cos(\phi_2)+2+c)$, with $c>0$. At small 
$\epsilon$, the KAM and Nekhoroshev theorems ([11],[12],[13]) apply to 
this kind of maps. The 
resonances of this system are defined by the straight lines 
$k_1 I_1+k_2 I_2 +2 \pi k_3=0$, with $k_1,k_2,k_3 \in {\Bbb Z}\backslash 0$, 
and the topology of the Arnold web on the section 
$S=\{(I_1,I_2)\in {\Bbb R}^2, \phi_i=0, i=1,2\}$ is that described for 
Hamiltonian system (1). 

\section{Global Arnold diffusion: numerical results}
A precise numerical detection of the Arnold web is possible with the so-called
Fast Lyapunov Indicator (hereafter, FLI) method [14]. 
In [15] we showed that 
 the Arnold web of system (1) indeed corresponds to 
the above theoretical description. In this paper, we are able to choose 
 initial conditions 
which are good candidates to diffuse
thanks to the accurate 
detection of the web provided in [15]. In fact, within resonances, 
both chaotic and regular motions are observed. Of course, regular 
motions do not diffuse, and therefore diffusive
orbits must be chosen in the small subset of the Arnold web made of chaotic 
motions. But even within chaotic motions, to have the chance to observe the 
phenomenon one has to restrict the choice of initial conditions to the 
single order resonances, i.e. to the portion of resonant lines which 
are far from the main crossing between resonances. The reason is that 
stability times are expected to be much longer at a distance of 
order $\sqrt{\epsilon}$ from these crossings (see, for example, [16]). 
With this selection of initial conditions, we have been able 
in [9] to show that indeed after times that increase 
faster than power law, initial conditions diffuse along 
the chaotic border of resonant lines. Though the computational time used in 
[9] was already quite long, due to the slowness of Arnold 
diffusion, it was not sufficient to evidence a global diffusion, i.e. 
the wandering of motions from one resonance to the others. In this 
paper we provide numerical experiments that thanks to a  much longer 
computational time, and proper choice of model parameters, display 
for the first time such global diffusion.  

The crucial parameters to set in the models are $\epsilon$ and the value of 
constant $c$ appearing in the denominator of the perturbation. In fact, 
any Fourier harmonic $\epsilon f_k=\epsilon \int f(\phi)
\exp(-i\sum_i k_i \phi){\rm d}\phi$ of the perturbing function $f$ is 
proportional  to $\epsilon$ and decreases (asymptotically) exponentially with 
$c\sum_i\norm{k_i}$. The values of the two 
parameters must be balanced so that $\epsilon$ is smaller than the critical 
value for Chirikov diffusion (determined with numerical methods, see 
[15], [17]) and the value of $c$ is not too large
so that harmonics with large order $\sum_i\norm{k_i}$
produce measurable effects on the finite time scale 
of our numerical computations. We found suitable values for our experiments 
$\epsilon=0.6$, $c=2$ for the map (2), and $\epsilon=0.01$, $c=1$ for the 
Hamiltonian system (1). The Arnold webs detected with the FLI method 
corresponding to these values are shown in figures 1,2. In the 
figures, for each point of the action plane $(I_1,I_2)$ we plot a color 
corresponding to the FLI value according to the color scale reported below 
the figure. 
The intermediate value of the FLI (orange in the color scale) corresponds to
KAM tori. Darker regions correspond to initial conditions $(I_1,I_2)$ 
on the section $S$ which produce resonant regular motions, 
while yellow regions correspond to initial conditions 
which produce chaotic motions. Each resonance
appears as a darker orange region with a yellow border, or as a single 
yellow line (depending on the resonance and on the value of the angles 
chosen to define the section $S$; see [15] for more details). Therefore, 
the yellow region on the FLI map corresponds to the chaotic subset of 
the Arnold web, where an eventual Arnold diffusion should be confined.
  
Following the explained criterion of choosing initial 
conditions within resonant chaotic motions far from the 
crossings of resonances we have chosen ten initial 
conditions near $(I_1,I_2)=(0.316,0.146)$ for the 
Hamiltonian case (see figure 2a), and twenty initial conditions near 
$(I_1,I_2)=(1.71,0.81)$ for the symplectic map (see figure 1a). Then, we 
computed numerically 
the map up to $10^{11}$ iterations, and we integrated numerically with a 
symplectic integrator the Hamiltonian system up to a $t\sim 10^{11}$ time 
units. 
The results are reported in figures 1 and 2: on the FLI map of the  
action plane $(I_1,I_2)$ we plotted as black dots all points of the orbits 
which have returned after some time on the section $S$. Of course, 
because computed orbits are discrete we 
represented points on the double section $|\phi _1| \leq 0.005$,   
$|\phi _2| \leq 0.005$ for both cases, and moreover $\phi _3 = 0$ for the 
Hamiltonian system; reducing the tolerance $0.005$ reduces only the 
number of points on the section, but does not change their diffusion 
properties.
Figures 
1a,2a show only the location of initial conditions (inside the circles), and 
figures 1b,2b show the result after the intermediate times 
$2. 10^9$ ($10^9$ for figure 2b). Figures 1c,2c,1d,2d show the result 
after much longer times. To properly display such long term evolutions we 
needed to use a zoomed out map of the action plane.   
\begin{figure}
\vskip -4 cm
 \hskip -2.6 cm~\includegraphics[height=26cm,width=20cm,angle=0]{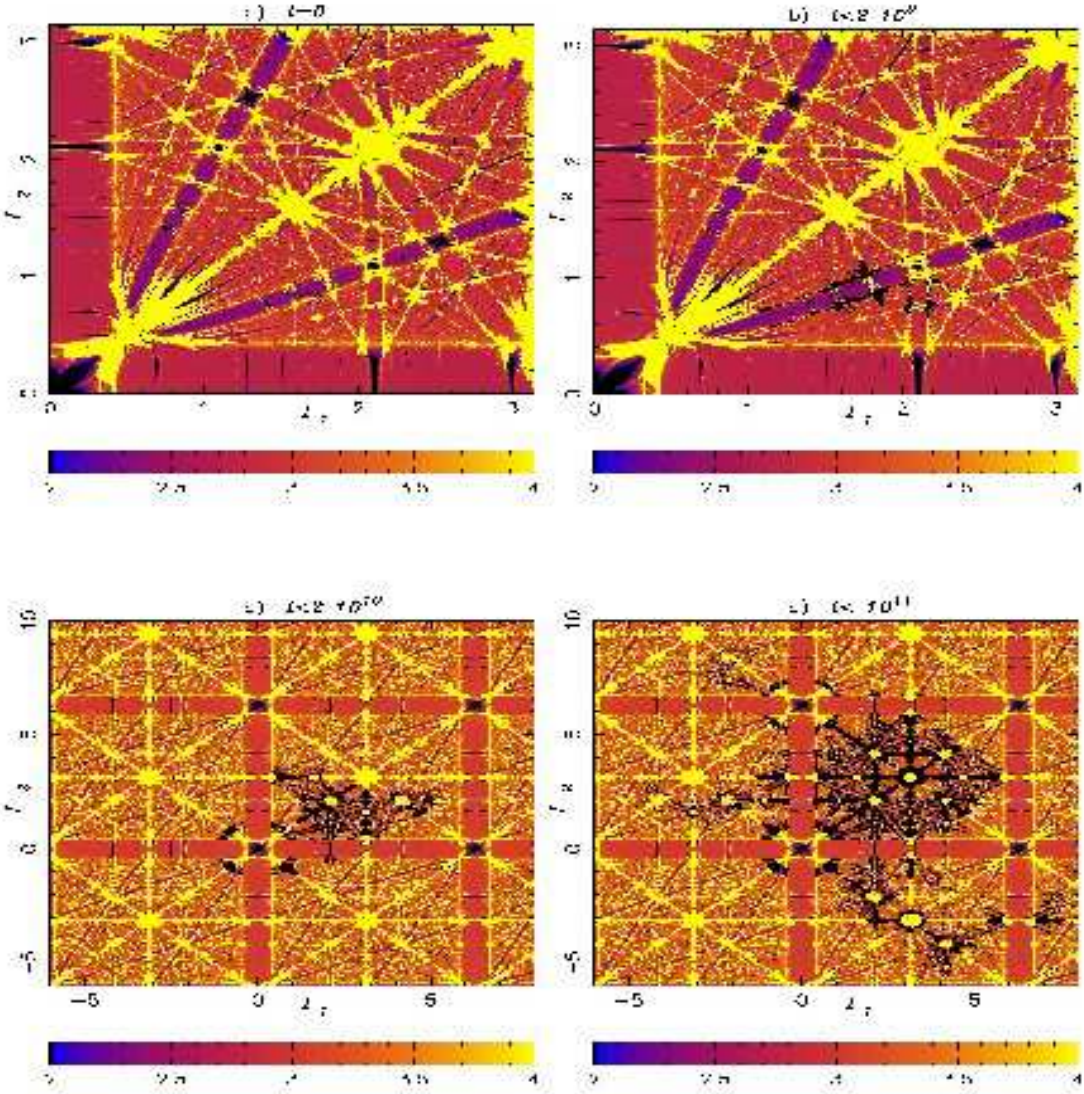}
\vskip -2 truecm
\caption{The four panels correspond to the FLI map of the 
action plane $(I_1,I_2)$ for the map (2), with initial condition on 
the section $S$ (see [15]), with different magnifications. The 
yellow region corresponds to the chaotic part of the Arnold web. 
Moreover, on panel (a) we mark with a circle the location of the twenty 
initial conditions; on panel (b,c,d) we mark with a black dot all 
points of the twenty orbits which have returned after some time on the 
section $S$.  We consider $2\,10^9$ iterations for panel (b); 
$2\,10^{10}$ iterations  for panel (c) and $10^{11}$ iterations for panel (d).
}
\end{figure}

\begin{figure}
\vskip -4 cm~\hskip -1 cm
\includegraphics[height=26cm,width=20cm,angle=0]{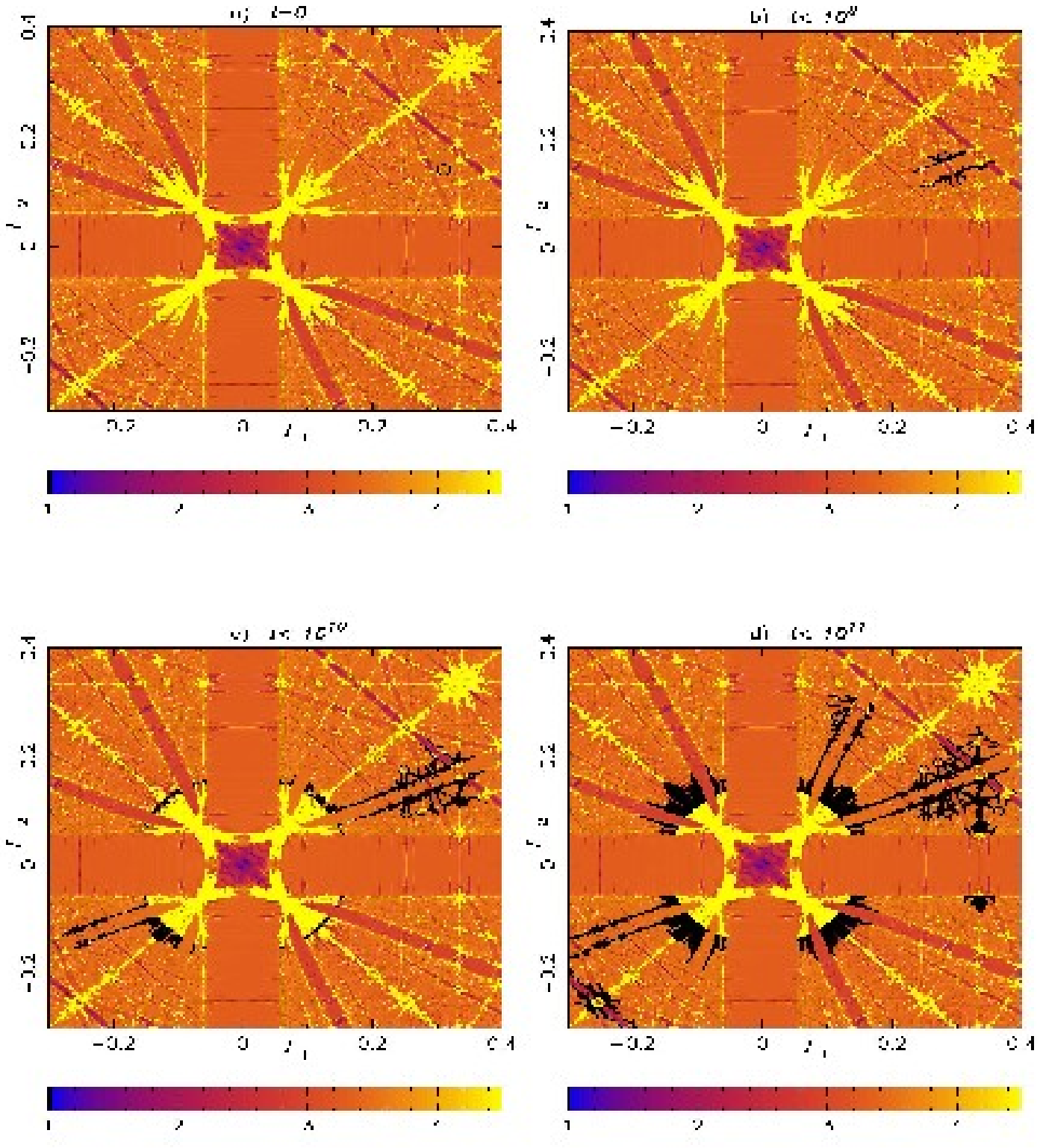}
\vskip -2 truecm
\caption{
The four panels correspond to the FLI map of the 
action plane $(I_1,I_2)$ for the Hamiltonian system (1), with initial 
condition on the section $S$ (see [15]), with different magnifications. 
The yellow region corresponds to the chaotic part of the Arnold web. 
Moreover, on panel (a) we mark with a circle the location of the ten 
initial conditions; on panel (b,c,d) we mark with a black dot all 
point of the twenty orbits which have returned after some time on the 
section $S$.  We consider $10^9$ iterations for panel (b); $1.2\ 10^{10}$ iterations  for panel (c) and $1.1\ 10^{11}$ iterations for panel (d).
}
\end{figure}

In both cases, the orbits filled a macroscopic region of the action plane
whose structure is clearly that of the Arnold web. 
The orbits have moved along the single resonances, and avoided the center of 
the main resonance crossings, in agreement with the  theoretical 
results which predict longer 
stability times for motions in these regions. The larger resonances 
(which correspond to the smallest orders $\vert k \vert$) are practically all 
visited, while this is not the case for the thinest ones (which correspond 
to the highest orders $\vert k \vert$). This is in agreement with the 
theoretical results of [18], which predict that the speed of 
diffusion on each resonance becomes smaller for resonances of high order. 
Therefore, the possibility of visiting all possible resonances is necessarily 
limited by finite computational time. 

On average, the drift behaves as a diffusion process. In fact, in the case of 
the map, the average evolution of the squared distance of $(I_1,I_2)$ from the 
initial datum, reported in fig. 3, increases almost linearly with time, 
so that we can measure a diffusion coefficient (the average slope of the plot) 
$D\sim 1.7\ 10^{-10}$ (we do not report the computation of the 
diffusion coefficient in the Hamiltonian case, which would require a longer 
computation time to explore a broader region of action space).

We remark that this diffusion coefficient characterizes the global diffusion 
process, while diffusion coefficients measured in [9] characterized 
local diffusion along a specific resonance. 
\begin{figure}
\includegraphics[height=10cm,width=8cm,angle=-90]{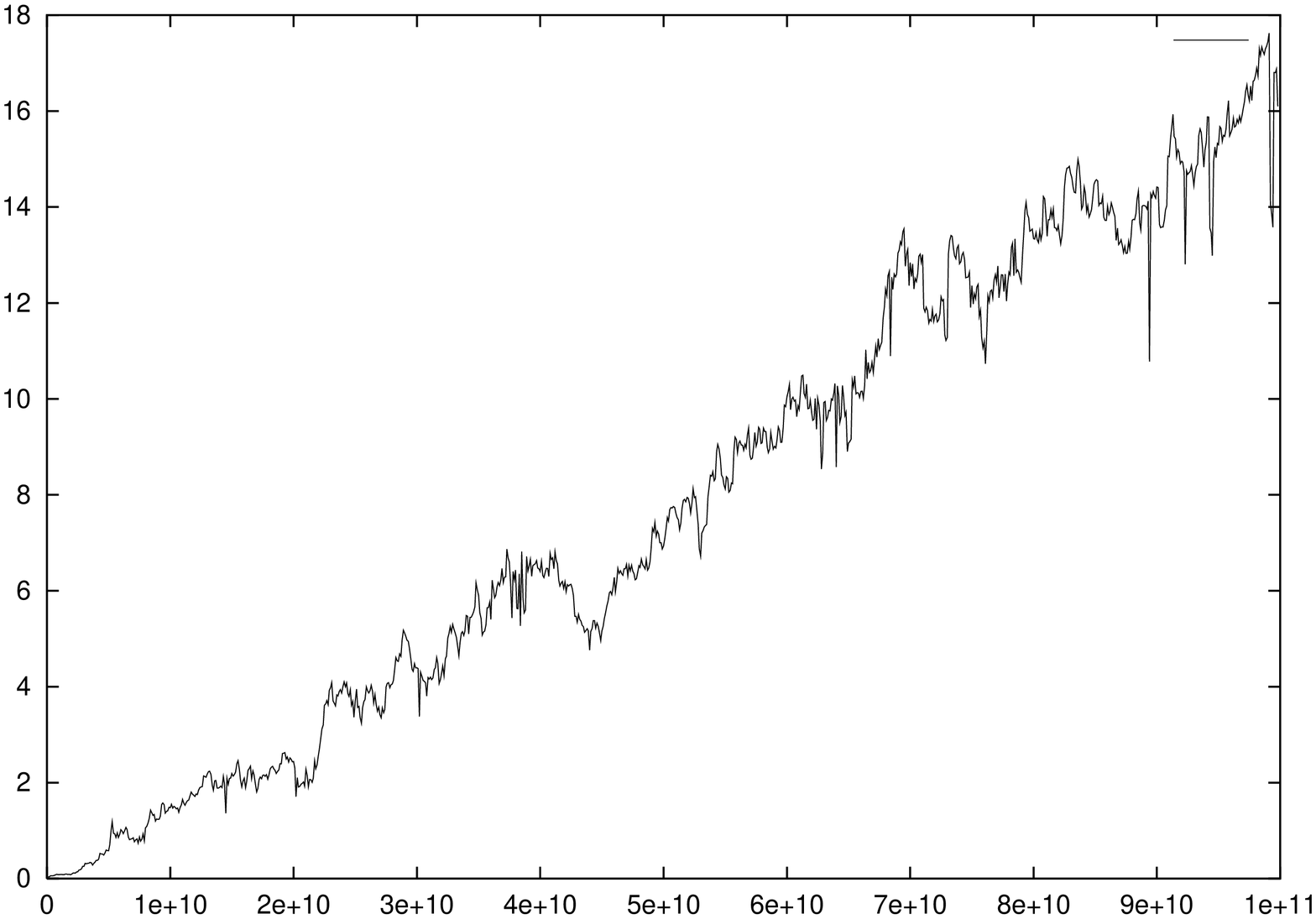}
\caption{Average evolution of the squared distance of $(I_1,I_2)$ from the 
initial datum for the map, measured for the points on the section $S$. The 
total computation time $t=10^{11}$ iterations has been divided in $10^3$ 
intervals. For each initial condition, and for each interval 
$[(n-1) 10^8,n 10^8]$, we have computed the average of the squared distance 
of $(I_1,I_2)$ from the initial datum, taking into account only points that 
in the interval $[(n-1) 10^8,n 10^8]$ are on the section $S$. Then, we 
averaged over all particles.} 
\end{figure}

\section{Conclusions}

In this paper we detect a macroscopic diffusion of orbits on the 
Arnold web of quasi--integrable Hamiltonian systems and symplectic maps.
The described diffusion phenomenon is very different from  
Chirikov diffusion, where the overlapping of resonances allows diffusion in 
macroscopic regions of phase space in relatively short time scales and 
without apparent peculiar topological properties of the stochastic region.  
The study of the long--term evolution in quasi--integrable systems, 
especially the set up of statistical methods, must therefore take into 
account that, at small perturbations, in a subset of phase space 
of peculiar structure there is an important phenomenon of diffusion of orbits. 
Indeed, it concerns various problems going from the old question 
of stability of the Solar System to the modern burden of the confinement of 
particles in accelerators.
\vfill
\eject

\noindent
{\bf References}\\
\vskip 0.4 truecm
\noindent
[1]  B.V. Chirikov. An universal instability of many dimensional
oscillator system. {\it Phys. Rep.}, {\bf 52}:263--379, (1979).\\
\noindent
[2]  A. Morbidelli. {\it Modern Celestial Mechanics. Aspects of Solar 
System Dynamics.} Taylor and Francis (2002).\\
\noindent
[3]  A.N. Kolmogorov. On the conservation of conditionally 
periodic motions under small perturbation of the Hamiltonian. 
 {\it Dokl. Akad. Nauk SSSR}, {\bf 98}:527, (1954). \\
\noindent
[4] J. Moser.   On invariant curves of area-preserving maps of an annulus.
{\it Comm. on Pure and Appl. Math.}, {\bf 11}:81--114, (1958).\\
\noindent
[5] V.I. Arnold. Proof of a theorem by A.N. Kolmogorov on the
invariance of quasi-periodic motions under small perturbations
of the Hamiltonian. 
{\it Russ. Math. Surveys.}, {\bf 18}:9--36, (1963).\\
\noindent
[6] N.N. Nekhoroshev. Exponential estimates of the stability 
time of near--integrable Hamiltonian systems.
{\it Russ. Math. Surveys}, {\bf 32}:1--65, (1977).\\
\noindent
[7] V.I. Arnold. Instability of dynamical systems with several 
degrees of freedom.  
{\it Sov. Math. Dokl.}, {\bf 6}:581--585, (1964).\\
\noindent
[8] J.~Mather. Announcement of result (2002).\\
\noindent
[9] E.~Lega, M.~Guzzo and C.~Froeschl\'e. Detection of Arnold
diffusion in Hamiltonian systems. 
{\it Physica D}, {\bf 182}:179--187 
(2003).\\
\noindent
[10] P.~Lochak. Canonical perturbation theory via simultaneous
approximations. {\it Russ. Math. Surv.}, {\bf 47}: 57--133.\\ 
\noindent
[11] S.~B.~Kuksin. On the inclusion of an almost integrable 
analytic symplectomorphism into a Hamiltonian flow.
{\it Russian Journal of Math. Phys.}, 
{\bf 1}:191--207 (1993).\\  
\noindent
[12] S.~B.~Kuksin and J.~P\"oschel. On the inclusion of 
analytic symplectic maps in analytic Hamiltonian flows and its 
applications.
{\it Nonlinear Differential 
Equations Appl.}, {\bf 12}: 96--116 (1994).\\
\noindent
[13] M.~Guzzo. A direct proof of the Nekhoroshev theorem for nearly integrable sysmplectic maps. To appear in Annales Henry Poincar\'e (2004).\\
\noindent
[14] C. Froeschl\'e, E. Lega and R. Gonczi. 
Fast Lyapunov Indicators. Application to Asteroidal Motion,
 Celest. Mech. and Dynam. Astron., {\bf 67},41-62,1997.\\
\noindent
[15] C.~Froeschl\'e, M.~Guzzo and E.~Lega. Graphical evolution of
the Arnold web: from order to chaos.
{\it Science}, {\bf 289}:2108--2110 
(2000).\\
\noindent
[16] J.~Poschel {\it Math. Z.}, 
Nekhoroshev estimates for quasi--convex hamiltonian systems.
{\bf 213}:187, (1993). \\
\noindent
[17] M.~Guzzo, E.~Lega and C.~Froeschl\'e. On the numerical 
detection of the effective stability of chaotic motions in 
quasi-integrable systems. 
{\it Physica D}, {\bf 163}:1--25 (2002).\\
\noindent
[18] A.~Morbidelli and A.~Giorgilli. On a connection between 
KAM and Nekhoroshev's theorems. {\it Physica D}, 
{\bf 86}:514--516 (1995).\\

\end{document}